# Realization of the SI Second Defined by Geometric Mean of Multiple Clock Transitions

Fang Fang[1,2*], Chaowei Wang[3], Yani Zuo[1,2], Shaoyang Dai[1,2]

[1] Time and Frequency Division, National Institute of Metrology, Beijing 100029, China
[2] Key Laboratory of Time Frequency and Gravity Primary Standard, State Administration for Market Regulation，Beijing 100029，China
[3] Department of Precision Machinery and Precision Instruments, School of Engineering Science, University of Science and Technology of China, Hefei, Anhui 230022, China

Email：fangf@nim.ac.cn

**Abstract**

The current definition of the SI second is based on the 133Cs ground-state hyperfine transition in the microwave domain, with the most accurate realizations achieving fractional frequency uncertainties of about $(1\text{-}2)\times10^{-16}$. In contrast, state-of-the-art optical clocks now demonstrate estimated uncertainties two to three orders of magnitude lower, prompting discussion on the redefinition of the SI second. Several options for the new definition have been proposed, one of which introduces a constant $N$ defined as the weighted geometric mean of multiple clock transition frequencies. In this work, we investigate how $N$ can be practically realized when not all defining transitions are available and when multiple optical clocks operate with different performance levels and non-overlapping uptimes. We consider two complementary realization and reconstruction routes. One route is based on geometric-mean combinations, and the other is based on arithmetic-mean combinations. We derive consistent uncertainty expressions that incorporate both measurement uncertainties and, where required, uncertainties of recommended frequencies or frequency ratios. Using analytic three-transition case studies, we identify the parameter regimes in which each route yields a lower total uncertainty and provide explicit conditions for the crossover between them. We further address the dominant role of dead time when a hydrogen maser serves as a flywheel reference by introducing a time-segmented, time-weighted combination based on coefficient and covariance matrices, which accounts for overlapping operation and correlations across measurement intervals. Our findings offer practical guidance for minimizing total uncertainty in multi-clock realizations and contribute to ongoing efforts toward redefining the SI second.

Keywords: optical clocks, SI second, frequency standard, weighted geometric mean, time scale calibration

## 1. Introduction

Time and frequency measurements are fundamental to a wide range of applications, including global navigation satellite systems [1,2], telecommunications [3,4], geodesy [5,6], and fundamental physics research [7-12]. Optical atomic clocks, leveraging ultra-narrow transitions, have now surpassed the stability and accuracy of caesium fountain clocks by more than two orders of magnitude [13-18]. This remarkable advancement has sparked discussions regarding the potential redefinition of the SI second based on optical transitions [19-21]. Currently, 11 optical transitions [22-32] are recognized as secondary representations of the second (SRS) by the Consultative Committee for Time and Frequency (CCTF), and their recommended values are updated regularly. Among these candidates, 5 species demonstrate evaluated Type B uncertainties of below $2\times10^{-18}$ [15-17,33,34], and 5 optical clocks have reached uncertainties in the $10^{-19}$ regime [15,17,34-36]. Notably, 4 distinct species have already contributed to the calibration of International Atomic Time (TAI), which is calculated by the Bureau International des Poids et Mesures (BIPM).

Despite these advancements, challenges remain in verifying clocks accuracy, improving operational uptime, and calibrating TAI with lower uncertainty to ensure its accuracy and conformity with the SI second [21,37].

After a series of discussions in CCTF, two options have been focused for the new definition. The first option extends the current single-transition approach and defines one chosen optical transition to be a fixed numerical constant. Reproducing this new definition would be similar to the current situation and straightforward, while other transitions would still be used as the secondary frequency standards for TAI calibration. The second option, first proposed by Lodewyck [38,39], defines the SI second by fixing the numerical value of the weighted geometric mean of the frequencies of a set of atomic clock transitions as a constant $N$. The numerical value of the constant $N$ and the weights are published by the CIPM, either as fixed values or updated according to the latest frequency ratio measurements. Under option 2, when only one clock transition is measured, the constant $N$ can be realized through its recommended frequency with an associate uncertainty, analogous to the SRS case. When multiple optical clocks are available, realizing the definition with minimum uncertainty becomes nontrivial, because not all defining transitions may be available in a single laboratory, clock performances can differ, measurements can be partially correlated, and operations are often unsynchronized with substantial dead time.Although extensive research has been conducted on optical clocks, discussions on how to realize the option 2 constant $N$ under such practical constraints remain limited. A key point is that multiple realizations and reconstruction strategies are possible. One strategy follows the multiplicative structure of the definition and forms a geometric-mean–based realization using measured frequencies (supplemented by recommended frequencies or frequency ratios when some transitions are not directly measured). Another strategy constructs an arithmetic-mean–type weighted combination of individual clock estimates. These approaches do not necessarily yield the same total uncertainty, and their relative performance depends on the balance between measurement uncertainties and recommended-value uncertainties, as well as on clock-to-clock performance mismatch and correlation structure.

This paper investigates the realization of the SI second using the composite-frequency approach (option 2) under various operational scenarios, with particular emphasis on minimizing total uncertainty. It provides consistent uncertainty expressions for both geometric-mean and arithmetic-mean realizations, and uses analytic three-transition case studies to identify the parameter regimes in which each method is advantageous and to give explicit crossover conditions. In addition, for asynchronously operated multi-clock campaigns referenced to a hydrogen maser, a time-segmented, time-weighted combination is introduced using coefficient and covariance matrices to account for overlapping operation, correlations, and dead-time–induced uncertainty propagation. Section 2 categorizes realization scenarios based on ensembles in which all transitions in the new definition are measured directly. Section 3 discusses situations where only a subset of the transitions in the new definition are measured directly and outlines the recommended frequency and associated uncertainties required to reproduce the definition. Section 4 explores real-world cases where multiple clocks operate with varying start time and durations. Section 5 concludes with perspectives on the potential future redefinition of the SI second.

## 2. Realization of the SI Second Using a Full Ensemble of Transitions

The proposed option 2 defines a constant $N$ based on an ensemble of transitions, as described in [38], as the weighted geometric mean of a set of clock transition frequencies, expressed by the following equation:

$$N = \prod_{i=1}^{n} v_i^{w_i}, \quad \sum_{i=1}^{n} w_i = 1. \tag{1}$$

where $v_i$ denote the recommended transition frequencies included in the new definition, and $w_i$ the normalized weights of each transition. It should be noted that both $N$ and $w_i$ are defining constants with no uncertainty, whereas $v_i$ represents an individual transition frequency with an associated uncertainty. The weights can be chosen according to certain rules [40,41] or determined by consensus.

### *2.1 Realization of the definition with arithmetical mean*

A straightforward way to reproduce the constant $N$ from an individual clock is given in Eq. 2.

$$\tilde{N}(i) = \tilde{v}_i \cdot \frac{N}{N_i} \quad (2)$$

Here $N_i$ is the recommended frequency of the *i*-th clock transition $v_i$ with a relative recommended uncertainty $u_{ii}$, and $\tilde{v}_i$ is the measured frequency of transition $v_i$ after correcting for biases induced by all relevant physical effects, with a relative uncertainty $u_i$. Therefore, the relative uncertainty of the reproduced $\tilde{N}(i)$ includes both the relative uncertainty $u_i$ of the measured frequency $\tilde{v}_i$ (incorporating Type A and Type B contributions), and the uncertainty $u_{ii}$ associated with the recommended $N_i$ as shown in Eq. 3.

$$u_{tot}^2(i) = u_i^2 + u_{ii}^2 \quad (3)$$

Each individual clock data can be used to measure the frequency of a timekeeping clock, as illustrated in Fig. 1, analogous to the use of a secondary frequency standard. The relative deviation of the timekeeping clock, $(f_{ref} - f_0)/f_0$, measured against the optical clock, can be obtained from Eq. 4.

$$\frac{f_{ref} - f_0}{f_0} = \frac{\tilde{N}(i) - N}{N} = \frac{\tilde{v}_i - N_i}{N_i} \quad (4)$$

This deviation can be used to steer the local time scale UTC(*k*) or to calibrate TAI.

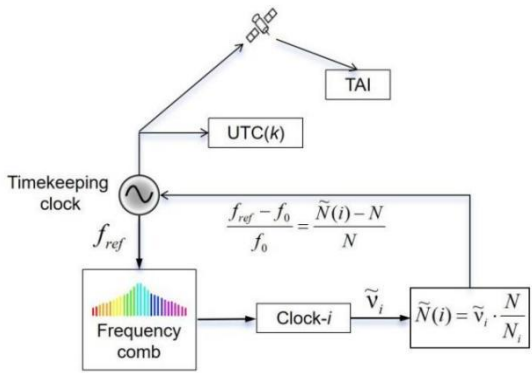

**Figure 1.** A schematic illustrating the reproduction of the constant *N* using a single optical transition measurement. The relative deviation of the timekeeping clock, $(f_{ref} - f_0)/f_0$, is measured against the optical clock frequency.

When all transitions are measured, the weighted arithmetic mean $\tilde{N}_{am}$, derived from Eq. 5, can be obtained and used to evaluate the deviation of the timekeeping clock.

$$\tilde{N}_{am} = \sum_i \left(\frac{u_{am}^2}{u_{tot}^2(i)}\right) \tilde{N}(i) \quad (5)$$

Here, $u_{am}$ is the total relative uncertainty of the weighted arithmetical mean, given by:

$$u_{am}^2 = \frac{\sum_i \left(\tilde{N}(i)/u_{tot}(i)\right)^2}{\left[\sum_i \tilde{N}(i)/u_{tot}^2(i)\right]^2} \quad (6)$$

When the reproduced $\tilde{N}(i)$ from different clocks are nearly equal, Eq. 6 can be approximated by Eq. 7 without compromising the accuracy of the result.

$$u_{am}^2 = \frac{1}{\sum_i 1/u_{tot}^2(i)} \quad (7)$$

*2.2 Realization of the definition with geometry mean*

An alternative way to reproduce the constant $N$ is through the geometric mean, as shown in Eq. 8.

$$\tilde{N}_{gm} = \prod_{i=1}^n \tilde{v}_i^{w_i} = \left(\prod_{i=1}^n k_i^{w_i}\right) \cdot f_{ref} \quad (8)$$

Each measured frequency $\tilde{v}_i$ is obtained by an optical frequency comb referenced to a timekeeping clock with an output frequency $f_{ref}$, such as a H-maser, and the ratio between them is denoted by $k_i$. The relative frequency bias of the timekeeping clock with respect to the defined constant $N$ can be obtained according to Eq. 9.

$$\frac{\tilde{N}_{gm} - N}{N} = \frac{\prod_i \tilde{v}_i^{w_i} - \prod_i N_i^{w_i}}{\prod_i N_i^{w_i}} = \frac{f_{ref} - f_0}{f_0} \quad (9)$$

Here, the $f_0$ is the nominal frequency of the timekeeping clock. When some optical clocks are referenced to different timekeeping clocks with an output frequency $f_i$ and the corresponding beat notes are measured, the reproduced $\tilde{N}_{gm}$ can be obtained using Eq. 10.

$$\tilde{N}_{gm} = \prod_{i=1}^n \tilde{v}_i^{w_i} \cdot \left(\frac{f_i}{f_1}\right)^{w_i} \quad (10)$$

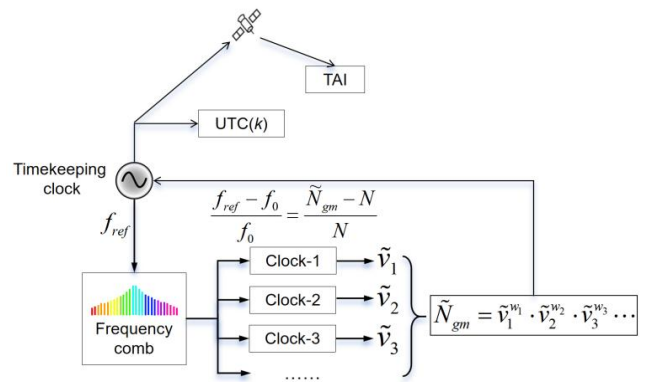

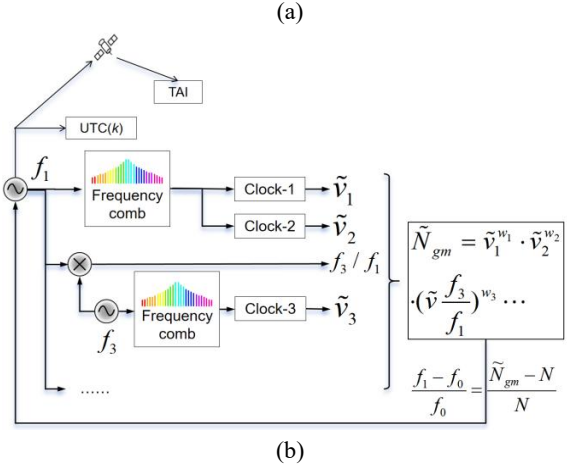

**Figure 2.** Schematic of reproducing the constant $N$ using optical transition measurements. (a) Realization of $N$ when all optical clocks are referenced to the same timekeeping clock. (b) Realization of $N$ when different optical clocks are referenced to different timekeeping clocks.

The geometric mean in Eq. 8 depends only on the measured frequencies, the recommended frequencies are not involved. Thus, the total uncertainty is determined solely by the uncertainties of the measured transition frequencies. In the simplest case, assuming that all clocks are referenced to the same timekeeping clock and that all measurements are uncorrelated, the total uncertainty is given by [42-44]:

$$u_{gm}^2 = \sum_i w_i^2 u_i^2 \qquad (11)$$

When difference references are used, the total relative uncertainty also includes the relative uncertainty of the measured frequency ratio between reference clock-$i$ and clock-1, denotes by $u_{i,1}$, as shown in Eq. 12.

$$u_{gm}^2 = \sum_i w_i^2 (u_i^2 + u_{i,1}^2) \qquad (12)$$

The uncertainty induced by dead time is more complicated and will be discussed in Section 4. For a normalized weighted mean over all transitions, the weights that minimize the total uncertainty are given by

$$w_i = \frac{1/u_{ii}^2}{\sum_k 1/u_{ii}^2} \qquad (13)$$

Thus, clocks with the lowest intrinsic uncertainty dominate the realization. One proposal for weight selection proposes that the weights be chosen inversely proportional to the square of the recommended clock transition uncertainty, as shown in Eq. 14.

$$w_i \propto 1/u_{ii}^2 \qquad (14)$$

In a specific case where the ratios $u_i/u_{ii}$ are identical for all clock transition measurements, Eq. 11 reduces to

$$u_{gm}^2 = \sum_i w_i^2 u_i^2 = \sum_i \left(\frac{1/u_i^2}{\sum_j 1/u_j^2}\right)^2 u_i^2 = \frac{1}{\sum_i 1/u_i^2} \qquad (15)$$

In practice, however, the measurements are often partially correlated, even for independent clocks located in different laboratories [43,44]. For example, thermometers used to measure temperature may have been calibrated against the same standard, leading to correlated uncertainties in blackbody-radiation-induced frequency shifts. In such cases, the total uncertainty in Eq. 15 must be replaced by Eq. 16:

$$u_{gm}^2 = \sum_i w_i^2 u_i^2 + 2\sum_{i=1}^{N-1}\sum_{j=i+1}^N w_i w_j u_i u_j r_{i,j} \qquad (16)$$

where $r_{i,j}$ is the correlation coefficient that quantifies the degree of correlation between two frequency measurements, ranging from $[-1, +1]$. Suppose that a physical quantity $x$ (such as temperature $T$, magnetic field $B$, etc.) introduces correlated contributions to the clock frequencies $v_i$ and $v_j$. The correlation coefficient $r_{i,j}$ can then be expressed as Eq.17,

$$r_{i,j} = \frac{\text{cov}(\tilde{v}_i, \tilde{v}_j)}{\sigma_i \sigma_j} \qquad (17)$$

Where $\sigma_i$ and $\sigma_j$ are the absolute uncertainties of $\tilde{v}_i$ and $\tilde{v}_j$, respectively. The covariance $\text{cov}(\tilde{v}_i, \tilde{v}_j)$ associated with the measured frequencies $\tilde{v}_i$ and $\tilde{v}_j$ is given, to first order, under the assumption that the measurement noises of different clocks are uncorrelated, by

$$\text{cov}(\tilde{v}_i, \tilde{v}_j) = \sum_{l=1}^L \frac{\partial F_i}{\partial x_l}\frac{\partial F_j}{\partial x_l} u^2(x_l) \qquad (18)$$

Here, $u(x_l)$ is the uncertainty of the parameter $x_l$. Without loss of generality, the measured frequency $\tilde{v}_i$ is understood as the directly observed value corrected for all frequency shifts induced by physical effects [42,47,48], and can be written as

$$\tilde{v}_i = \tilde{v}_i^{\text{meas}} - F_i(x_1, x_2, \cdots) \qquad (19)$$

where $x_i$ are physical quantities that introduce frequency biases $F_i$, such as temperature, laser intensity, or atomic density.

## 2.3 Case Study: Realization with Three Transitions

To illustrate the dependence of the total uncertainty on these parameters, the simplest case is considered when $N$ is the geometric mean of three optical transitions whose recommended values share the same relative uncertainty $u_{11}$, and therefore have equal weights. The reproduced geometric mean and its uncertainty is expressed in Eq. 20 when all measurements have the same relative uncertainty $u$,

$$\tilde{N}_{gm} = \tilde{v}_1^{1/3} \cdot \tilde{v}_2^{1/3} \cdot \tilde{v}_3^{1/3}$$
$$u_{gm}^2 = \frac{u^2}{3} \quad (20)$$

While the arithmetic mean and its relative uncertainty are given by

$$\tilde{N}_{am} = \frac{N}{3} \sum_i \frac{\tilde{v}_i}{N_i}$$
$$u_{am}^2 = \frac{u^2 + u_{11}^2}{3} \quad (21)$$

Compared with Eq. 20, the total uncertainty obtained from the arithmetic mean includes the recommended uncertainty $u_{11}$ and is therefore worse. More generally, different clocks may exhibit measurement uncertainties that deviate from the recommended uncertainty $u_{11}$ by different amounts. In a special case that two optical clocks have the same measurement uncertainty $u$ and that the third one has a measurement uncertainty $ku$. The geometric mean remains the same as in Eq. 20, whereas the arithmetic mean is given by Eq. 22:

$$\tilde{N}_{am} = \frac{\frac{1}{u^2+u_{11}^2}}{\frac{2}{u^2+u_{11}^2}+\frac{1}{k^2u^2+u_{11}^2}}\left(\tilde{v}_1 \cdot \frac{N}{N_1} + \tilde{v}_2 \cdot \frac{N}{N_2}\right)$$
$$+ \frac{\frac{1}{k^2u^2+u_{11}^2}}{\frac{2}{u^2+u_{11}^2}+\frac{1}{k^2u^2+u_{11}^2}} \tilde{v}_3 \cdot \frac{N}{N_3} \quad (22)$$

Again, assuming that all the measurements are uncorrelated, the total uncertainties obtained with the geometric-mean and arithmetic-mean methods can be obtained from Eq. 23:

$$u_{gm}^2 = \frac{2}{9}u^2 + \frac{1}{9}(ku)^2 = \frac{2+k^2}{9}u^2$$
$$u_{am}^2 = \frac{1}{2/(u^2+u_{11}^2)+1/(k^2u^2+u_{11}^2)} \quad (23)$$

The comparison of the uncertainties in the realization of the constant $N$ using the arithmetic mean and the geometric mean under different conditions is shown in Figure 3. It indicates that when the measurement uncertainty $u$ lies within a certain range, the total uncertainty of the geometric mean $u_{gm}$ is smaller than the uncertainty of the arithmetic mean $u_{am}$, provided that the measured uncertainty is relatively low or the ratio $k$ falls within a specific interval (not far from 1). The total uncertainty of the geometrical mean increases remarkably as $k$ increases, and at certain points, $u_{gm}$ exceeds $u_{am}$. The intersection point depends on the ratios of $k$ and $u/u_{11}$, as shown in Fig. 3.

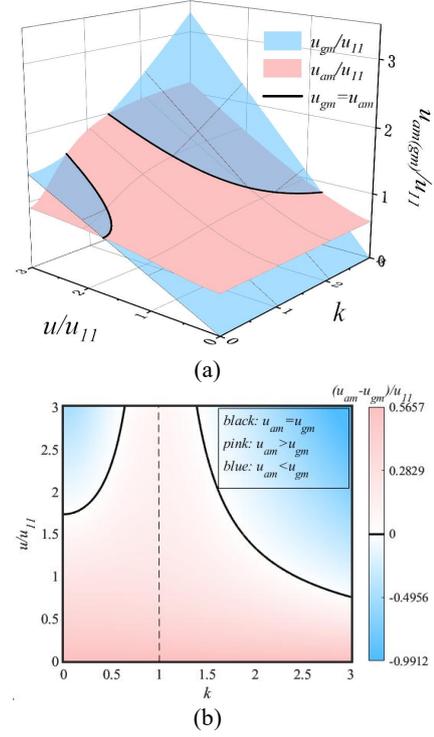

(a)

(b)

**Figure 3.** The total uncertainties of realization of the constant $N$ using two different methods. (a) The blue plane represents the total uncertainty in realizing $N$ using the geometric mean of three optical clocks, while the pink plane represents the total uncertainty using the arithmetic mean. The black line of intersection denotes the condition where the uncertainties from the two methods are equal. Here, two clocks have the same uncertainty $u$, while the third one has uncertainty $ku$. The recommended uncertainties $u_{11}$ are assumed to be the same for all three transitions. (b) Projection of the intersection line of the two surfaces in (a), indicating the difference of the uncertainties obtained from the geometric-mean and arithmetic-mean methods.

Next, a more general scenario is considered in which $N$ is the geometric mean of three optical transitions. Two of these transitions have the same recommended uncertainties uncertainty $u_{11}$ and the same weight $m$, whereas the third is distinct, as given in Eq. 24.

$$N = v_1^m \cdot v_2^m \cdot v_3^{1-2m} \quad (24)$$

Similarly, the reproduced $N$ obtained from geometric-mean and arithmetic-mean methods can be expressed by Eq. 25 when the measured transition $\tilde{v}_1$ and $\tilde{v}_2$ have the same uncertainty $u$ and $\tilde{v}_3$ has an uncertainty $ku$:

$$\tilde{N}_{gm} = \tilde{v}_1^m \cdot \tilde{v}_2^m \cdot \tilde{v}_3^{1-2m}$$

$$\tilde{N}_{am} = \frac{\dfrac{u^2+u_{11}^2}{2}}{\dfrac{2}{u^2+u_{11}^2}+\dfrac{1}{k^2u^2+\dfrac{m}{1-2m}u_{11}^2}} \left( \tilde{v}_1 \cdot \frac{N}{N_1} + \tilde{v}_2 \cdot \frac{N}{N_2} \right) \quad (25)$$

$$+ \frac{\dfrac{1}{k^2u^2+\dfrac{m}{1-2m}u_{11}^2}}{\dfrac{2}{u^2+u_{11}^2}+\dfrac{1}{k^2u^2+\dfrac{m}{1-2m}u_{11}^2}} \tilde{v}_3 \cdot \frac{N}{N_3}$$

Again, assuming uncorrelated measurements, the total uncertainties for realization of the constant $N$ using the two methods can be obtained accordingly. In a case that the ratio of the recommended uncertainty of the third clock to that of the first two is assumed to match the ratio of their weights:

$$\frac{m}{1-2m} = \left(\frac{u_{33}}{u_{11}}\right)^2 \quad (26)$$

$$u_{gm}^2 = 2m^2 u^2 + (1-2m)^2 (ku)^2$$

$$u_{am}^2 = \frac{1}{\dfrac{2}{u^2+u_{11}^2}+\dfrac{1}{k^2u^2+\dfrac{m}{1-2m}u_{11}^2}} \quad (27)$$

Figure 4 illustrates that, similar to Fig. 3, when the measurement uncertainty $u$ is relatively low or ratio of the uncertainty of the third clock to that of the other two clocks lies within a certain interval (when $k$ is close to $\sqrt{m/(1-2m)}$) the total uncertainty obtained from the geometric mean is smaller than that derived from the arithmetic mean. As the performance of the third clock deteriorates (as $k$ increases), the geometric-mean uncertainty first increases slowly and then rises rapidly, whereas the arithmetic-mean uncertainty increases rapidly at first and then grows more slowly, with the line of intersection indicates where the uncertainties associated with the two methods are equal, and consistent with the results shown in Fig. 3. Furthermore, the larger $m$ is, the wider the range over which the geometric mean provides an advantage.

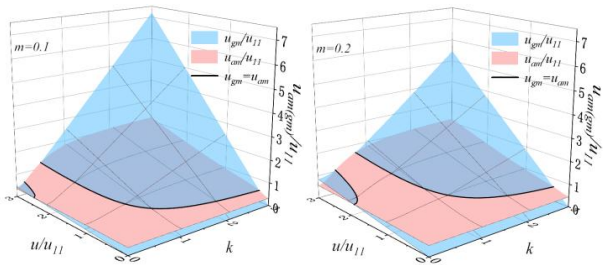

**Figure 4.** The total uncertainties of realization of the constant $N$ using two different methods. (a) The blue plane represent the ratio $u_{gm}/u_{11}$, while the pink plane shows the ratio $u_{am}/u_{11}$. The black intersection line indicates the condition where the uncertainties from the two methods are equal. Here, two clocks have the same measured uncertainty $u$, and the third has uncertainty $ku$. The weights of the first two clocks are same and denoted as $m$, while the weight of the third clock is $1-2m$. (b) Variation of the dominance regions where the for the two uncertainties as the value of $m$ changes, where geometric mean having a lower uncertainty compared to that of the arithmetical mean method. The area enclosed by two curves of the same color indicates the region in which the geometric mean gives a lower uncertainty.ss

## 3. Realization of the SI Second Using a Subset of Transitions

### 3.1 Different methods for realization of the definition

It is not feasible to measure every transition included in the definition within a single laboratory. Nevertheless, the composite constant $N$ can still be reconstructed using either the geometric-mean or the arithmetic-mean approach, based on measured data and recommended frequencies. It is straightforward to obtain an arithmetic mean $\tilde{N}_{am}$ from the measured $\tilde{v}_i$ from Eq. 5. The only difference here is that not all transitions are included. The total uncertainty can also be calculated accordingly using Eq. 7.

To obtain $N$ via the geometric mean, the recommended values $N_j$ must be used for transitions that are not measured directly. To ensure that the relative bias of the calculated $\tilde{N}_{gm}$ is proportional to the relative bias of the reference time keeping clock signal, and thus suitable for calibrating time scale, frequency ratio $N_j/N_1$ should be used rather than absolute frequencies. Here, transition 1 is selected among the measured transition with the smallest measured uncertainty.

Then $\tilde{N}_{gm}$ is calculated using Eq. 28:

$$\tilde{N}_{gm} = \tilde{v}_1^{w_1+\sum_j w_j} \cdot \prod_{i\neq 1,j} \tilde{v}_i^{w_1} \cdot \prod_j \left(\frac{N_j}{N_1}\right)^{w_j} \quad (28)$$

where $\tilde{v}_1$ is a measured transition with the lowest uncertainty, $\tilde{v}_i$ denote the other measured transition frequencies, $N_j$ are the recommended frequencies for the unmeasured transitions, and $N_1$ is the recommended frequency for transition 1.

The relative bias with respect to the defined constant $N$ can be expressed as Eq. 29,

$$\frac{\tilde{N}_{gm}-N}{N} = \frac{\tilde{v}_1^{w_1+\sum_j w_j}}{N_1^{w_1+\sum_j w_j}} \cdot \prod_{i\neq 1,j}\frac{\tilde{v}_i^{w_i}}{N_i^{w_i}} - 1 = \frac{f_{\text{ref}}-f_0}{f_0} \quad (29)$$

The associated uncertainty of $\tilde{N}_{gm}$ is

$$u_{gm}^2 = (w_1+\sum_j w_j)^2 u_1^2 + \sum_{i\neq j,1} w_i^2 u_i^2 + \sum_j w_j^2 u_{j1}^2 \quad (30)$$

Here, $u_{j1}$ is the recommended uncertainty of the frequency ratio $N_j/N_1$. When only one transition is measured, Eq. 28 reduces to

$$\tilde{N}_{gm} = \tilde{v}_1 \cdot \prod_{j\neq 1}\left(\frac{N_j}{N_1}\right)^{w_j} = \tilde{v}_1 \cdot \frac{N}{N_1} \quad (31)$$

which is identical to Eq. 2, reproducing $N$ through the ratio between $N$ and the recommended $N_i$. If some measurements are correlated, as in Eq. 16, the total uncertainty becomes

$$u_{gm}^2 = (w_1+\sum_j w_j)^2 u_1^2 + \sum_{i\neq j,1} w_i^2 u_i^2 \\ + 2\sum_{i=1}\sum_{k=i+1} w_i w_k u_i u_k r_{i,k} + \sum_j w_j^2 u_{j1}^2 \quad (32)$$

*3.2 Case Study: Realization with Three Transitions*

For a concrete example, again consider $N$ defined as the geometric mean of three optical transitions first, each with the same recommended uncertainty. Suppose that two transitions are measured and the third uses its recommended frequency. The realization of $N$ through geometrical mean is,

$$\tilde{N}_{gm} = \tilde{v}_1^{2/3} \cdot \tilde{v}_2^{1/3} \left(\frac{N_3}{N_1}\right)^{1/3} \quad (33)$$

and the total uncertainty becomes

$$u_{gm}^2 = (\frac{2}{3})^2 u_1^2 + (\frac{1}{3})^2 u_2^2 + (\frac{1}{3})^2 u_{31}^2 \quad (34)$$

The realization of $N$ through the weighted arithmetic mean yields

$$\tilde{N}_{am} = \frac{\frac{1}{u_1^2+u_{11}^2}}{\frac{1}{u_1^2+u_{11}^2}+\frac{1}{u_2^2+u_{22}^2}}\tilde{v}_1 \cdot \frac{N}{N_1} \\ + \frac{\frac{1}{u_1^2+u_{11}^2}}{\frac{1}{u_1^2+u_{11}^2}+\frac{1}{u_2^2+u_{22}^2}}\tilde{v}_2 \cdot \frac{N}{N_2} \quad (35)$$

With the corresponding uncertainty

$$u_{am}^2 = \frac{1}{\frac{1}{u_1^2+u_{11}^2}+\frac{1}{u_2^2+u_{22}^2}} = \frac{1}{\frac{1}{u_1^2+\frac{1}{3}u_{31}^2}+\frac{1}{u_2^2+\frac{1}{3}u_{31}^2}} \quad (36)$$

Here, $u_{11} = u_{22} = \frac{\sqrt{3}}{3}u_{31}$ as calculated in the reference[38].

A comparison of the total uncertainties produced by the two different methods is shown in Fig. 5. The x- and y-axes represent the ratios of the measured uncertainties to the recommended uncertainty (i.e., $a = u_1/u_{11}$ and $b = u_2/u_{11}$ with $a \leq b$), while the z-axis gives the ratio of the total realization uncertainty to the recommended uncertainty $u_{11}$. The blue plane represent the ratio $u_{gm}/u_{11}$, while the pink plane shows the ratio $u_{am}/u_{11}$.

The result demonstrates that the arithmetic-mean method generally yields a lower total uncertainty when one transition has a significantly larger measured uncertainty, whereas the weighted geometric-mean method has an advantage when $\sqrt{2}u_1 - \frac{\sqrt{3}}{2}u_{11} < u_2 < \sqrt{2}u_1 + \frac{\sqrt{3}}{2}u_{11}$. when the measurement uncertainties of the clocks are smaller than twice the recommended uncertainty (i.e., $u_1/u_{11} < 2$ and $u_2/u_{11} < 2$), the geometric-mean method almost always yields a lower total uncertainty.

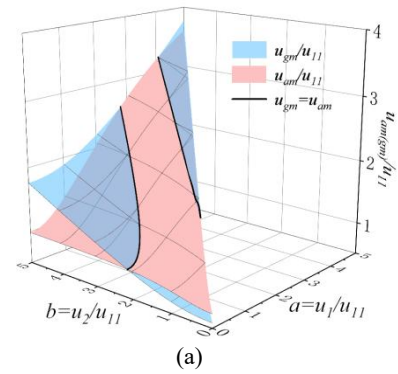

(a)

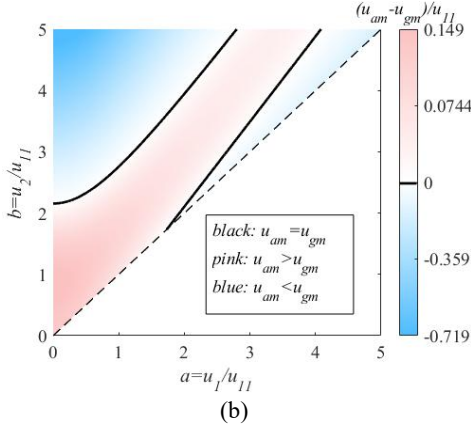

(b)

**Figure 5.** Comparison of the total uncertainties of realization of N with two different methods. (a) The horizontal axes represent the normalized measurement uncertainties $a = u_1 / u_{11}$ and $b = u_2 / u_{11}$. The blue plane indicates the ratio $u_{gm} / u_{11}$, while the pink plane indicates the ratio $u_{am} / u_{11}$. The intersection line of the two planes marks the condition where the uncertainties from the two methods are equal. (b) Projection of the intersection line from (a).

## 4. Total Uncertainty in the Unsynchronized Operation of Multiple Clocks

When a hydrogen maser is used as the reference signal, dead time during optical clock operation leads to a significant contribution to the measurement uncertainty [49,50]. For example, for an uptime of 80% (corresponding to 20% dead time), the dead-time-induced uncertainty $u_{lab}$ reaches approximately $3.6 \times 10^{-16}$, which is much larger than the intrinsic Type A and Type B uncertainties of a state-of-the-art optical clock [49].

Figure 6 illustrates the additional uncertainty introduced by different dead-time fractions for a hydrogen maser with a fractional frequency stability of $2.0 \times 10^{-16}$.

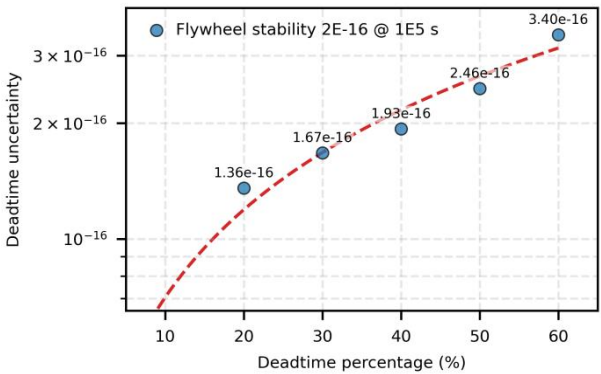

**Figure 6.** Additional uncertainty introduced by different dead-time fractions for a hydrogen maser with a fractional frequency stability of $2.0 \times 10^{-16}$.

As a consequence, the total uncertainty $u_i$ of each optical clock in a campaign of duration $T$ can become large if only the mean value over the entire campaign is used. To reduce this contribution when several optical clocks are operated asynchronously, we divide the campaign into multiple measurement intervals and form a weighted combination,

$$\tilde{N}_{Tam} = \sum_i \frac{t_i}{T} \tilde{N}(t_i) \quad (37)$$

The assignment of measurement intervals is illustrated in Fig. 7, which shows an example of four optical clocks operating with different uptimes during a campaign of duration $T$.

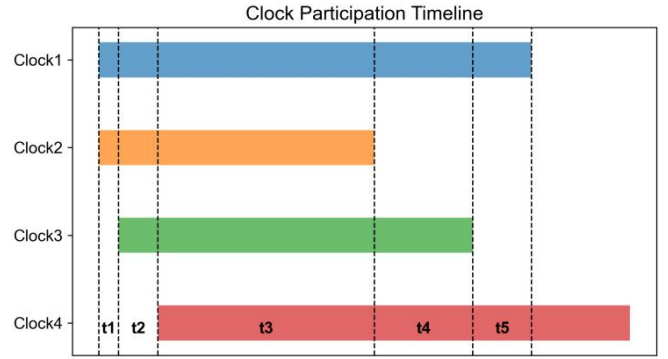

**Figure 7.** Example of four optical clocks operating at different times during a campaign of total duration $T$.

For each interval $t_i$, the measured frequency and its uncertainty are determined as described above. Then, similar to Eq. 7, when the realization of $N$ from different clocks in different time periods are about the same, the total uncertainty of the combined final result can be obtained from Eq. 38,

$$u_T^2 = C^T U C \quad (38)$$

where $C$ is the coefficient matrix defining the weighted linear combination $\left(\frac{t_1}{T}, \frac{t_2}{T}, \cdots, \frac{t_n}{T}\right)^T$, and $U$ is the covariance matrix. The matrix $U$ is an $n \times n$ matrix with elements:

$$u(i,i) = \sum_{p \neq q}^{n} [\alpha_{i,p} \alpha_{i,q} \frac{1}{\tilde{v}_{i,p} \tilde{v}_{i,q}} \sum_{l=1}^{L} \frac{\partial F_p}{\partial x_l} \frac{\partial F_q}{\partial x_l} u^2(x_l)] + \sum_{p=1}^{n} \alpha_{i,p}^2 u_{i,p}^2$$

$$u(i,j) = \sum_{p=1}^{n} [\alpha_{i,p} \alpha_{j,p} \frac{1}{\tilde{v}_{i,p} \tilde{v}_{j,p}} \sum_{l=1}^{L} \frac{\partial F_p}{\partial x_l} \frac{\partial F_q}{\partial x_l} u^2(x_l)] + \sum_{p=1}^{n} \beta_{i,p} \beta_{j,p} u_{pp}^2$$

(39)

Here, $\tilde{v}_{i,p}$ is the measurement value of clock $p$ during the $i$-th period, $F_p$ is the frequency shift correction for clock $p$, $u_{i,p}$ is the relative uncertainty of the measurement value of clock $p$ during the $i$-th period, and $u_{pp}$ is the relative uncertainty of the recommended value for clock $p$. If the same clock contributes to both intervals $i$ and $j$, the corresponding uncertainties are fully correlated. The coefficients $\alpha_{i,p}$ and $\beta_{i,p}$ are determined as follows:

$$\alpha_{i,p} = \begin{cases} w_{p_i} + W_i^{\text{miss}}, & p = p_i \\ 0, & p \in A_i \\ w_p, & p \in S_i. \end{cases} \quad (40)$$

$$\beta_{i,p} = \begin{cases} -W_i^{\text{miss}}, & p = p_i \\ w_p, & p \in A_i \\ 0, & p \in S_i. \end{cases} \quad (41)$$

where $p_i$ denotes the index of the clock with the smallest relative measurement uncertainty in the i-th period, $A_i$ is the set of indices of all non-operational clocks during this period, $S$ represents the set of indices of operational clocks in the same period, excluding clock $p_i$, and $W_i^{\text{miss}}$ is the total weight of all non-operational clocks in the i-th period.

By adopting this framework, the impact of maser-induced dead time is minimized, and the effective uncertainty of the composite measurement can be substantially reduced compared with treating each clock independently.

## 5. Conclusions

This paper focuses on the realization of the SI second using Option 2, with an emphasis on uncertainty evaluation under various operational scenarios. The uncertainties of realization of the constant *N* with geometric and arithmetic mean methods were derived, and the contributions from the measurement uncertainties of the involved optical transitions, the recommended frequency uncertainties and frequency ratio uncertainties are analyzed. The different scenarios including all clock transitions and only a subset of transitions being available are considered, and correlations between measurements are also considered. Special cases with three transitions in different scenarios are studies, show in which scenario the geometric mean is more appropriate.

To mitigate dead-time-induced uncertainty, a time-weighted-mean approach was introduced, using coefficient and covariance matrices to account for overlapping operations and correlations among clocks. These results highlight the practical advantage of Option 2–style weighted combinations for reproducing the SI second. The proposed framework provides a systematic method for combining frequency measurements from asynchronously operated optical clocks and segmented measurement campaigns, explicitly incorporating maser-flywheel dead-time–induced uncertainty and correlations effects via a coefficient/covariance-matrix formalism. Its applicability to arbitrary numbers of clocks and measurement segments makes it directly relevant to SI second realization workflows and to evaluation/steering tasks such as UTC(k) operation and TAI calibration.

Future work may include incorporating correlations between different reference standards, developing adaptive weighting schemes, and extending the methodology to optical clock networks across laboratories. Overall, this approach supports multi-clock SI second realizations and TAI calibrations while minimizing dead-time-induced uncertainty under real-world operational constraints.

## Acknowledgements

This work is supported by the National Key Research and Development Program of China (Grant No. 2021YFF0603801), the Basic Research Expenses Program of National Institute of Metrology China (NO. AKYZD2411).